\documentclass[prb,twocolumn,showpacs, superscriptaddress]{revtex4}
\usepackage{graphicx}
\usepackage{amsmath}
\usepackage{amsxtra}
\usepackage{amssymb}
\usepackage{dcolumn}
\usepackage{float}
\usepackage{bm}
\usepackage[breaklinks=true,colorlinks,citecolor=blue,linkcolor=blue,urlcolor=blue]{hyperref}

\def\nn{\nonumber}
\def\bc{\begin{equation}}
\def\ec{\end{equation}}
\def\bea{\begin{eqnarray}}
\def\eea{\end{eqnarray}}

\begin{document}

\title{Anisotropic plasmons, Friedel oscillations and screening  in $8-Pmmn$ borophene}
\author{Krishanu Sadhukan}
\affiliation{Department of Physics, Indian Institute of Technology, Kanpur 208016, India}
\author{Amit Agarwal}
\email{amitag@iitk.ac.in}
\affiliation{Department of Physics, Indian Institute of Technology, Kanpur 208016, India}

\date{\today}

\begin{abstract}
$8-Pmmn$ borophene is a polymorph of borophene which hosts an anisotropic tilted Dirac cone. Using a low energy effective Hamiltonian for borophene, we analytically calculate the density-density response function of borophene and study its anisotropic plasmon dispersion and screening properties. We find that the anisotropic plasmon mode in $8-Pmmn$ borophene, remains undamped for higher energies along the mirror symmetry direction. The friedel oscillation in borophene is found to decay as $r^{-3}$ in the large $r$ limit, with both the amplitude as well as the periodicity of the Friedel oscillations being anisotropic. 
\end{abstract}

\maketitle

\section{Introduction}

Collective density oscillations (plasmons) of an interacting electron gas, and their screening properties, are of fundamental interest from both experimental  and theoretical perspective \cite{Polini_review,Maier,Stauber,GV}. From an application point of view, plasmons in graphene and related two dimensional (2D) materials are very promising on account of their longer lifetime, gate tunability and optoelectronic applications -- facilitating the localization and guiding of light into electrical signals\cite{appl}. 
Plasmons and the dynamical dielectric function have been actively explored in a variety of 2D materials including spin-orbit coupled 2D electron gas \cite{fabian_2010,SOC_plasmon,Persistent}, 2D hole gas \cite{scholz_2012}, graphene \cite{stauber_2006,sarma_2007,polini_2011,amit_2015,Polini_review,Peres}, MoS$_2$, \cite{scholz_2013}, topological insulators \cite{TI_plasmons}, massive Dirac materials \cite{pyat_2009,thakur_2015,thakur_2017}, and phosphorene \cite{TLow1, Rodin_plasmon, Barun_plasmon}.    
In this paper, we study the plasmon dispersion, dynamical dielectric function and static screening properties in one of the 2D polymorphs of borophene \cite{Poly1, Poly2,Feng,Mannix}, which has a symmetry protected tilted and anisotropic Dirac cone \cite{zabo_2016}. 

The polymorph of borophene with a tilted and anisotropic Dirac cone (called $8-Pmmn$ borophene) was predicted in Ref.~[\onlinecite{Zhou1}], and has recently been experimentally confirmed \cite{Feng2}. 
The {\it ab-initio} properties of $8-Pmmn$ borophene monolayer have been  studied in detail\cite{lopez_2016}, and based on symmetry considerations an effective low energy Hamiltonian, for energies and wave-vectors in vicinity of the Dirac point (laying on the $\Gamma-Y$ axis), was proposed in Ref.~[\onlinecite{zabo_2016}].  Using the low energy effective Hamiltonian of $8-Pmmn$ borophene, we analytically calculate the density-density response function (also called the polarization/Lindhard function) and the dynamical dielectric function within the random phase approximation (RPA). We 
show that $8-Pmmn$ hosts anisotropic plasmons, and has highly anisotropic static screening properties in the long wavelength screening as well as the Friedel oscillations, which decays as $r^{-3}$ far away from the impurity.  

This article is organized as follows: In Sec.~\ref{sec2} we use the anisotropic 2D Dirac Hamiltonian of borophene to analytically calculate the density density response function. In Sec.~\ref{sec3}, we use the analytical polarization function to study the anisotropic plasmon dispersion. This is followed by a detailed discussion of the long wavelength screening and anisotropic Friedel oscillations in Borophene, in Sec.~\ref{sec4}. 
In Sec.~\ref{sec5} we explore the impact of  small disorder on the plasmon dispersion of borophene. Finally we summarize our findings in Sec.~\ref{sec6}. 

\section{Non-interacting Polarization Function for $8-Pmmn$ Borophene}
\label{sec2}
Our starting point is the effective low energy Hamiltonian of $8-Pmmn$ borophene, which describes a tilted anisotropic Dirac cone in 2D using three parameters \cite{zabo_2016}: 
\bc \label{eqH}
\mathcal{H}= \hbar \sum_{\bf k} \psi_{\bf k}^\dagger \left[v_x k_x \sigma_x + v_y k_y \sigma_y + v_t k_y \openone \right] \psi_{\bf k}~.
\ec
Here $\psi_{\bf k}=(a_{\bf k} , b_{\bf k})^T$, where $a_{\bf k}$ and $b_{\bf k}$ are the the destruction operators of the Bloch states of the two triangular sub-lattices in $8-Pmmn$ borophene. 
The three velocities are given by \cite{zabo_2016} $\{v_x,v_y,v_t\} = \{0.86, 0.69, 0.32\} \times10^6$ m/s. The corresponding energy dispersion in 
given by $\varepsilon_{s{\bf k}} = \hbar v_t k_y + s \hbar \sqrt{v_x^2k_x^2 + v_y^2k_y^2}$, and $s=+1$ ($-1$) denotes the conduction (valance) band. Similar to the crystal structure of $8-Pmmn$ borophene (see Ref.~[\onlinecite{lopez_2016,zabo_2016}]), its energy dispersion also has mirror symmetry about the ${\hat y}$ axis (${\hat y}-{\hat z}$ mirror plane at $x=0$). This mirror symmetry will also reflect in the polarization function, plasmon dispersion, and the static screening properties discussed below. 

The non-interacting dynamical polarization function of an electron gas with two bands, for a given (${\bf q},\omega$) is given by 
\bea \label{response_func}
\Pi^{(0)}&=&\frac{g}{4\pi^2} \sum_{{\bf k}, s, s'}\frac{n_{{s\bf k}}-n_{{s'\bf k+q}}}{\hbar\omega^+ + \varepsilon_{s{\bf k}}- \varepsilon_{s'{\bf k+q}}}
f_{ss'}({\bf k},{\bf q}),
\eea
where $g = 2$ denotes the spin ($=2$) and valley ($=1$) degeneracy,  $\omega^+ = \omega + i \eta$ with $\eta \to 0$, and  $f_{ss'}({\bf k},{\bf q})$ is the orbital overlap function. The Fermi function is given by $n_{s{\bf k}}=(e^{\beta(\varepsilon_{s{\bf k}} - \mu)} +1)^{-1}$, where $\beta = 1/(k_B T)$ and $\mu$ denotes the chemical potential. At zero temperature the Fermi function acts as a step function and it sets the integration limits. 
In what follows we will work in the $\omega>0$ regime, since we have $\Pi^{(0)}({\bf q},-\omega) = \Pi^{(0)} ({\bf q},\omega)^*$. 

\begin{figure}
\includegraphics[width=1.0\linewidth]{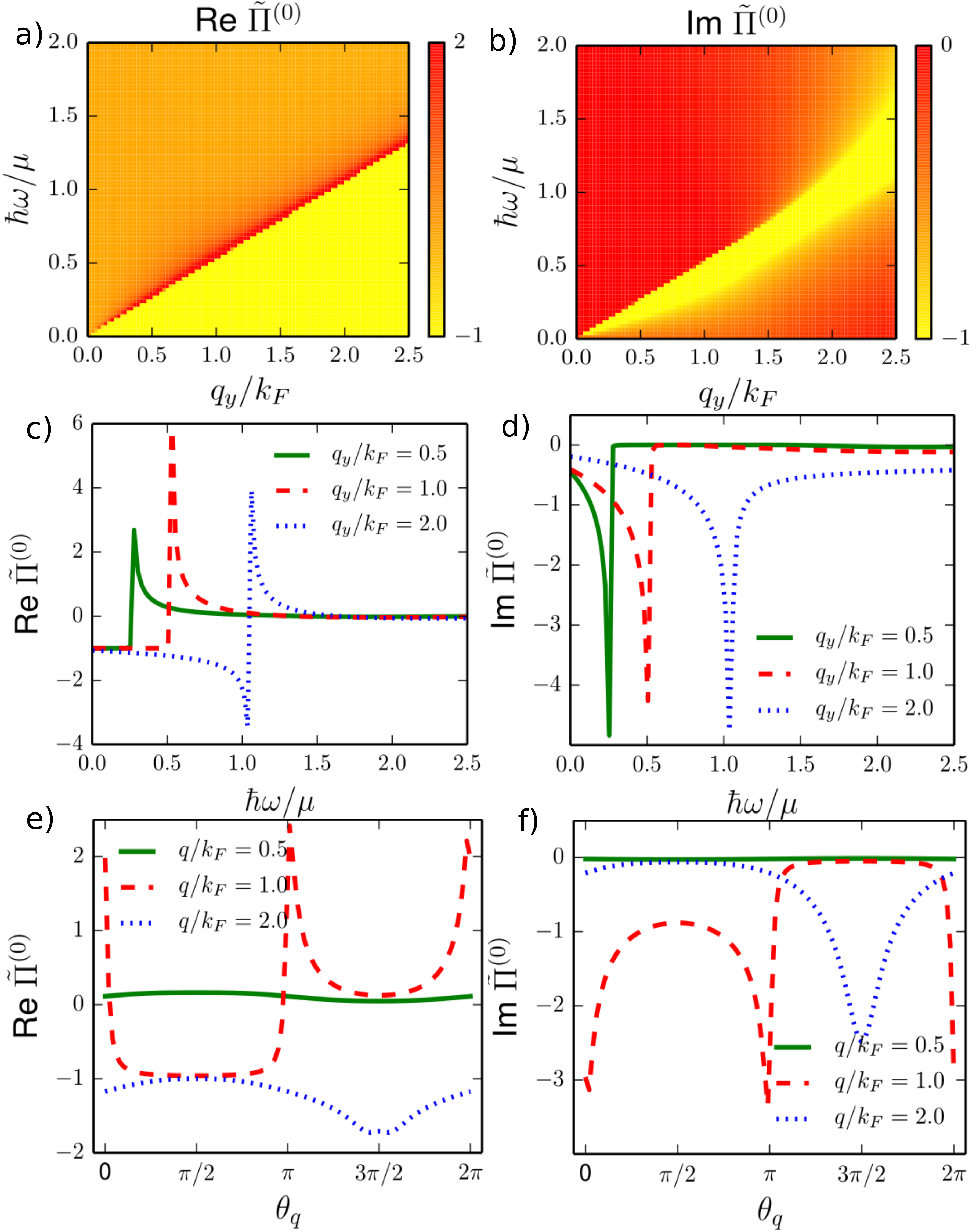}
\caption{(a) Real part and (b) imaginary part of the polarization function. Here $\bf q$ is taken along $y$ axis ($q_x =0$). (c) Real part and (d) imaginary part of the polarization function as a function of $\omega$ for fixed $q$ values. (e) Real part and (f) imaginary part of the polarization function as a function of $\theta_q$ which is the angle between $x$ axis and $\bf q$ vector. In panels e) and f) we have chosen $\hbar \omega = \mu$, and   the three $q$ values have been chosen such that, $E_q < \hbar \omega $, $E_q=\hbar \omega$ and $E_q> \hbar \omega$.  In all the panels the polarization functions are in units of $N_0 = {g\mu}/({2\pi\hbar^2v_xv_y})$, and we have defined $k_F\equiv {\mu}/{\hbar v_x}$. 
\label{fig1}} 
\end{figure}

To simplify the calculation of the anisotropic polarization function, we define the following modified variables: 
\bea
E_{kx}=\hbar v_{x} k_{x}, ~~~~ E_{ky}=\hbar v_{y} k_{y}, ~~{\rm and}~~ E_k = \sqrt{E_{kx}^2+E_{ky}^2}~.\nn \\
\eea
In terms of the new variables the energy eigenvalue is $\varepsilon_{s{\bf k}}=\hbar v_t k_y + s E_k$.  
The band overlap function, in terms of the redefined variables, is given by 
\bc
f_{ss'}({\bf k},{\bf q})=\frac{1}{2}\left(1+ss'\frac{E_k+E_q \cos \phi_E}{\sqrt{E_k^2+E_q^2+2 E_k E_q \cos\phi_E}}\right),
\ec
where $\phi_E=\tan^{-1}({E_{ky}}/{E_{kx}})-\tan^{-1}({E_{qy}}/{E_{qx}})$. 

The non-interacting polarization function in Eq.~\eqref{response_func}, can be expressed as a sum of the polarization function of the intrinsic part $\Pi^{(0)}_0$ (which is finite for $\mu = 0$), and an extrinsic part $\Pi^{(0)}_1$ (which vanishes for $\mu \to 0$): $\Pi^{(0)} = \Pi^{(0)}_0 + \Pi^{(0)}_1$. Both the intrinsic and the extrinsic contributions are evaluated explicitly in appendix \ref{PC}, in a manner similar to that in graphene\cite{pyat_2009}.
The total polarization function can  be expressed in a compact form as  
\bc \label{Pi_final}
\frac{\Pi^{(0)}({\bf q},\omega)}{N_0} = -1+\frac{1}{8 \mu}\frac{E_q^2}{\sqrt{E_q^2-\hbar^2\omega'^2}}
\big[\Gamma^+(\nu^+)+ \Gamma^-(\nu^-)\big]~.
\ec
Here $N_0\equiv{g\mu}/({2\pi\hbar^2v_xv_y})$ is the density of states in 2D anisotropic massless Dirac systems, 
and we have defined $\omega' \equiv \omega - v_t q \sin\theta$. In Eq.~\eqref{Pi_final} we have defined the complex function 
\bc
\Gamma^{\pm}(z) \equiv z\sqrt{1-z^2}\pm i\cosh^{-1}(z)~,
\ec  
and the dimensionless variables, 
\bc
\nu^{\pm}({\bf q},\omega) \equiv \frac{2\mu\pm\hbar(\omega-v_tq_y)}{E_q}~.
\ec  

In Fig.~\ref{fig1} (a)-(b), we display the real and the imaginary part of the polarization function in the $\omega - q_y$ plane. Since 
the anisotropic term in the $8-Pmmn$ borophene Hamiltonian [see Eq.~\eqref{eqH}] involves $q_y$ only, thus along the $q_x$ direction (for $q_y =0$) the polarization function resembles that of graphene. Panels (c) and (d) of Fig.~\ref{fig1} show the real and imaginary part of the polarization function respectively as a function of $\omega$ for different $q_y$ values, while panels  (e) and (f) explore the same as a function of the 
orientation angle of $\bf q$ in the $q_x-q_y$ plane for a fixed value of $\omega$ and $q$. Note the mirror symmetry of the polarization function about the ${\hat y}$ axis, {\it i.e.}, about the $\theta_q = \pi/2$ and $\theta_q = 3\pi/2$ line in Fig.~\ref{fig1}(e)-(f). No such symmetry exists along the $\hat x$ axis or alternately around $\theta_q = 0$ and $\theta_q = \pi$ lines in Fig.~\ref{fig1}(e)-(f).
In the next section, we use the obtained polarization function to calculate the dispersion of the dynamical collective density  excitations,  {\it i.e.}, plasmons. 

\section{Plasmon Dispersion}
\label{sec3}
The dynamical interacting density-density response function, within the random phase approximation (RPA), is given by 
\bc \label{Piint}
\Pi^{\rm RPA}({\bf q},\omega) = \frac{\Pi^{(0)}({\bf q},\omega)}{1-v_q \Pi^{(0)}({\bf q},\omega)}~,
\ec
where $v_q = 2 \pi e^2/(\kappa q)$ is the Coulomb interaction in the momentum space in 2D with $\kappa$ being the static dielectric constant. Collective density excitations or plasmons (within  RPA) are now given by the poles of $\Pi^{\rm RPA}({\bf q},\omega)$, or alternately by the zeros of the dynamical dielectric function $\epsilon({\bf q},\omega)$, {\it i.e.},  
\bc\label{plasmon}
\epsilon({\bf q},\omega)\equiv 1-v_q \Pi^{(0)}({\bf q},\omega)=0~.
\ec

\begin{figure}[t]
\includegraphics[width=\linewidth]{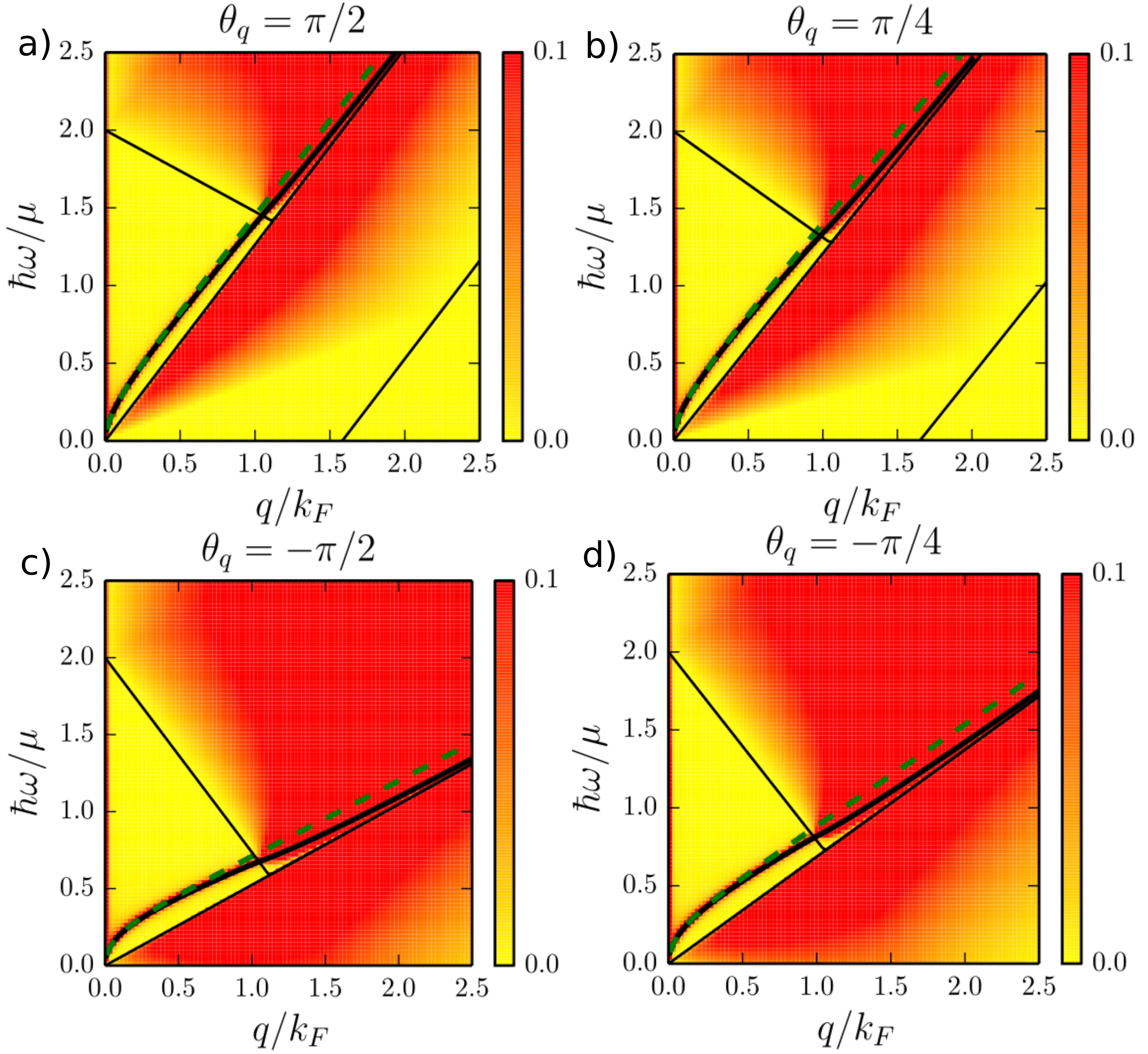}
\caption{The anisotropic loss function ${\cal{E}}_{\rm loss}({\bf q},\omega) =-\mathfrak{Im}[1/\epsilon({\bf q},\omega)] $ in borophene along with the numerical (solid black) and the analytical (dashed green) - Eq.~\eqref{dispersion} - plasmon dispersion curve  for (a) $\theta_q = \pi/2$ (b) $\theta_q = \pi/4$  (c) $\theta_q = -\pi/2$ and (d) $\theta_q = -\pi/4$. In all the panels the thin black lines denote the particle-hole continuum boundaries specified by Eqs.~\eqref{b1}-\eqref{b4}. We have chosen $2 \pi e^2 N_0/\kappa k_F =q_{\rm TF}/k_F = 4/\pi$ for all the panels.  
\label{fig2}} 
\end{figure}

From an experimental perspective, plasmons appear as resonance peaks in the momentum resolved electron energy loss spectrum, which directly measures the loss function \cite{Barun_plasmon}: ${\cal{E}}_{\rm loss}({\bf q},\omega) =-\mathfrak{Im}[1/\epsilon({\bf q},\omega)] $. The anisotropic loss function for $8-Pmmn$ borophene is shown in Fig.~\ref{fig2} for $\theta_q = \pi/2$, $\pi/4$, $-\pi/2$ and $-\pi/4$ in panels a), b) c) and d) respectively.  In all the panels, there is a dominant plasmon peak in the region without any (single) particle-hole excitations, and it broadens (gets damped) significantly on entering the inter-band particle hole continuum -- whose boundary is also $\theta_q$ dependent. 

The particle hole boundary for the intra-band transitions is given by lines specified by 
\bc \label{b1}
\omega^{\rm intra}_+({\bf q}) = v_x q \left[f(\theta_q) + v_t \sin\theta_q/v_x\right]~,
\ec 
and 
\bc \label{b2}
\omega^{\rm intra}_-({\bf q}) = v_x q \left[f(\theta_q) + v_t \sin\theta_q/v_x\right] - 2 \mu/\hbar~,
\ec
where we have defined the function 
\bc \label{b3}
f^2(\theta_q)=\cos^2\theta_q+\frac{v_y}{v_x}\sin^2\theta_q~.
\ec
In terms $f(\theta_q)$ we have  $E_q=\hbar v_x q {f(\theta_q)}$.  
Similarly the particle hole boundary of the inter-band single particle excitations is given by 
\bc \label{b4}
\omega^{\rm inter} = 2 \mu /\hbar - v_x q [f(\theta_q) - v_t \sin\theta_q/v_x]~. 
\ec 
For $\theta_q = \pi/2$ ($\theta_q = -\pi/2$), both the boundaries of the intra-band and the inter-band particle hole continuum are pushed to higher (lower) energies for a given $q$. Now if the plasmon mode decays by entering into the inter-band particle hole continuum, then it remains undamped for relatively higher (lower) energies for $\theta_q = \pi/2$ (for $\theta_q =-\pi/2$) -- as shown in panel (a) (panel (c)) of Fig.~\ref{fig2}.

To obtain the exact plasmon dispersion, Eq.~\eqref{plasmon} has to be solved numerically. However analytical expressions can be obtained in the long wavelength limit. In the {\it dynamical long wavelength limit} with ${\bf q} \to 0$ and $\omega> v_x q$, the polarization function can be approximated as\cite{amit_2015}, 
\bc\label{long}
\Pi^{(0)}({\bf q},\omega)\approx -N_0\bigg[1-\frac{\omega'}{\sqrt{\omega'^2 -v_x^2q^2f^2(\theta_q)}}\bigg]~.
\ec
Using Eq.~\eqref{long} in Eq.~\eqref{plasmon},  the long wavelength plasmon dispersion can be obtained to be  
\bc\label{dispersion}
\omega_{\rm pl}(q\to 0) = v_x q f(\theta_q)\frac{1+v_qN_0}{\sqrt{1+2v_qN_0}}+v_tq\sin\theta_q ~. 
\ec
The second term in the r.h.s of Eq. \eqref{dispersion} is linear in $q$ and the co-efficient becomes negative for $\theta_q \in (0,-\pi)$. Thus it seems that the long wavelength plasmon frequency first increases and then decreases with increasing $q$ as the linear term becomes dominant. However, this turns out to be a peculiarity of the long wavelength limit, and the actual (exact) plasmon dispersion remains monotonically increasing and eventually decays by entering the inter-band particle hole continuum. This is highlighted in Fig.~\ref{fig2}, which displays the exact plasmon dispersion (solid black line) along with the corresponding long wavelength limit (green-dashed line), on the backdrop of the momentum resolved loss function for different values of $\theta_q$. 

\section{Static Response and Charged impurity screening}
\label{sec4}

\begin{figure}
\includegraphics[width=\linewidth]{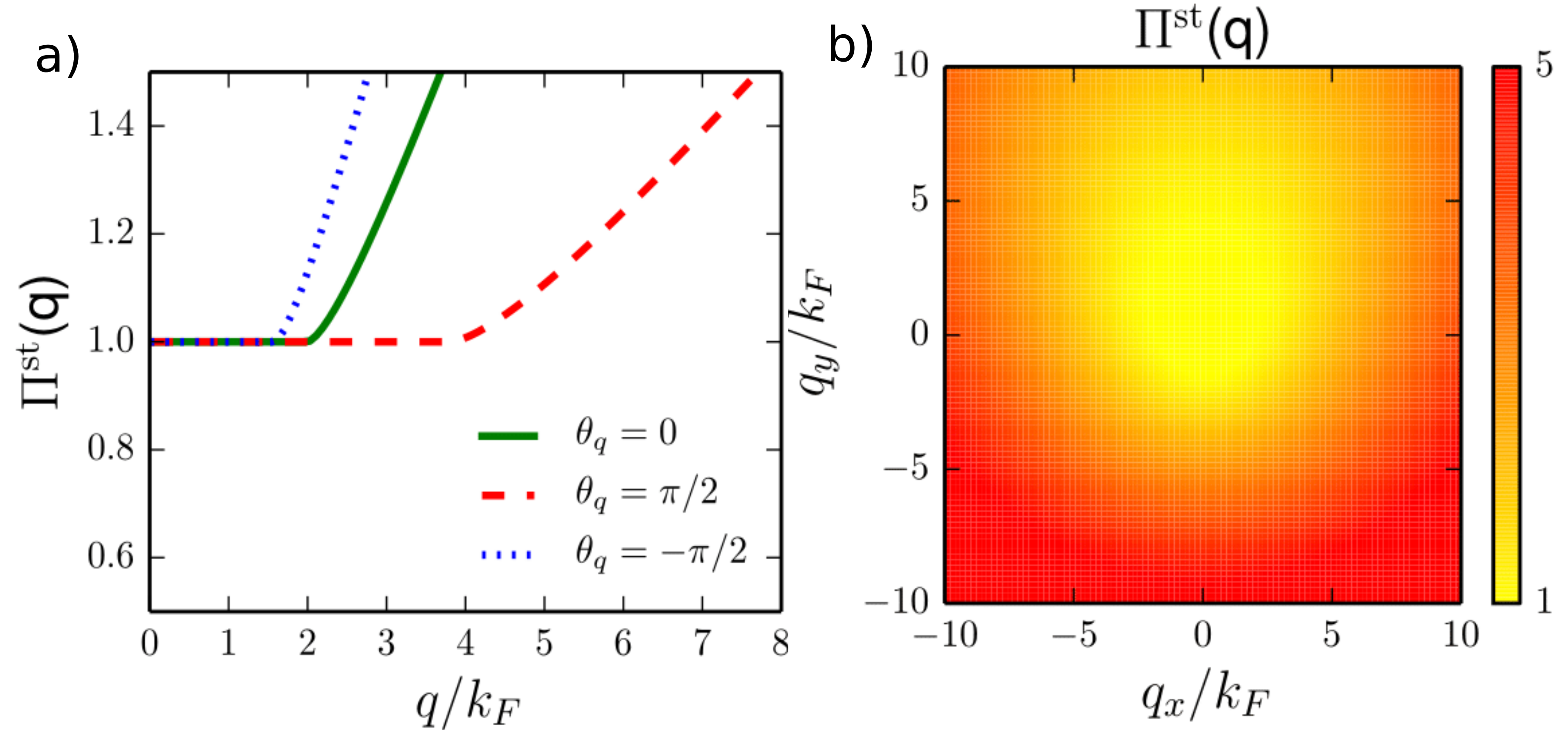}
\caption{(a) The anisotropic static response function (in units of $\Pi^{\rm st}(0) = -N_0$) as a function of $q$, for different $\theta_q$. 
For $q< q_0(\theta_q)$, $\Pi^{\rm st}({\bf q}) = N_0$, and for $q>q_0(\theta_q)$ it shows anisotropic behaviour. 
(b) Color plot $\Pi^{\rm st}({\bf q})$ in the $q_x-q_y$ plane. Note that it has a mirror symmetry about the y-axis ($q_x=0$ line), while there is no such symmetry 
along the x-axis ($q_y = 0$ line) .
\label{fig3}}
\end{figure}

We now turn our attention to the static scenario ($\omega \to 0$) which is useful for studying the long wavelength screening of coulomb interactions in an electron gas, and the corresponding Friedel oscillations of a charged impurity. The static non-interacting response function is given by $\Pi^{\rm st}({\bf q}) \equiv \Pi^{(0)}({\bf q},0)$ and 
\bea\label{static}
\frac{\Pi^{\rm st}({\bf q})}{N_0}&=&\Bigg[-1+\Theta\left[q{f(\theta_q)}-q v_t \sin\theta_q/v_x-2k_F\right]\\
&\times& \frac{q{f(\theta_q)}}{4k_F\sqrt{1-\frac{v_t^2\sin^2\theta_q}{v_x^2f^2(\theta_q)}}} G_<\left(\frac{2k_F+q v_t \sin\theta_q/v_x}{q{f(\theta_q)}}\right)
 \Bigg],\nn~ 
\eea
where $k_F = {\mu}/{\hbar v_x}$ and we have defined a real valued function\cite{stauber_2006} $G_<(x) \equiv x\sqrt{1-x^2}-\cos^{-1}(x)$. 
In the long wavelength limit, or more generally for $q < q_0$ where 
\begin{equation} \label{q0}
q_0 (\theta_q) \equiv \frac{2 k_F}{f(\theta_q) - v_t \sin(\theta_q)/v_x}~,
\end{equation} 
we have $\Pi^{(0)}( q< q_0,0) = -N_0$,  a constant. However for finite $q> q_0$ the static dielectric function is anisotropic. The anisotropic static response function is shown in Fig.~\ref{fig3}(a), for $q$ along three different directions. Figure ~\ref{fig3}(b) shows the static response function in the $q_y-q_x$ plane, where the mirror symmetry about the ${\hat y}$ axis ($q_x=0$) is evident. 

The effective screened potential of a charged impurity at origin, of potential $\phi_{\rm ext}(\bf q)$ in vacuum, is given by 
\bc\label{screen}
\phi_{\rm sc}({\bf r})=\int d{\bf q}e^{i{\bf q}.{\bf r}}\frac{\phi_{\rm ext}(\bf q)}{\epsilon({\bf q},0)}e^{-qs}~.
\ec
Here $s\rightarrow0$ is supposedly the distance of the charged impurity from the 2D plane of the electron gas. We now discuss the long wavelength Thomas-Fermi screening, and 
the Friedel oscillations of a charged impurity in the next two subsections. 

\subsection{Long wavelength Thomas Fermi screening}
For large length scales or in the long wavelength ($q\to 0$) limit, the static screening of a charged impurity in an electron liquid is dictated by the long wavelength static dielectric constant. 
Using Eq.~\eqref{static}, we have 
\begin{equation} \label{eps_long}
\epsilon(q < q_0,0) = 1 + q_{\rm TF}/q~, 
\end{equation}
where we have defined a density (or $\mu$ dependent) Thomas Fermi wavevector, $q_{\rm TF} \equiv 2 \pi e^2 N_0/\kappa =  g e^2 \mu/(\hbar^2 \kappa v_x v_y) $ in all directions. The $q_{\rm TF}$ for $8-Pmmn$ borophene is almost identical to that of graphene with $v_{\rm F}^2 \to v_x v_y$, barring the extra valley degeneracy factor in graphene. The Thomas Fermi screening wavevector is arising specifically from the extrinsic contribution of the static polarization function which vanishes for the intrinsic case ($\mu \to 0$). 
The long wavelength screening in 8mmn borophene reduces the impact of a  Coulomb impurity (with $\phi_{\rm ext}({\bf r}) = e^2/r$), and its effective behaviour decreases as $r^{-3}$ for large distances ($r > 1/q_{\rm TF}$), similar to that in graphene\cite{stauber_2006}. 

\subsection{Anisotropic Friedel oscillations}
We now discuss the impact of screening of a charged impurity by the electron gas in $8-mmn$ borophene arising from the dielectric contribution for finite wavevectors. This typically leads to Friedel oscillations in the system with a power law decay \cite{thakur_2017}. 

In the asymptotic $r \to \infty$ limit, the integral in Eq.~\eqref{screen} has a very rapidly varying phase: $qr \cos(\theta_q - \theta_r)$, where $\theta_r$ is the the real space angle denoting the direction of ${\bf r}$. Consequently  the dominant contribution to it arises when the phase  is stationary. Now following Ref.~[\onlinecite{fabian_2010}] we use the method of stationary phase which allows us to simplify Eq.~\eqref{screen}. The $\theta_q$ dependence of $\epsilon({\bf q},0)$ in Eq.~\eqref{screen} is replaced by $\theta_q \to \theta_r$. With the $\theta_q$ dependence of $\epsilon({\bf q},0)$ taken care of, the angular integration \cite{simion_2005} in Eq.~\eqref{screen} can be easily performed to yield,
\bc \label{eqS}
\int_0^{2\pi} e^{iqr\cos\theta_q} d\theta_q=2\pi J_0(qr)\simeq\frac{\sqrt{8\pi}}{\sqrt{qr}}\cos(qr-\pi/4)~.
\ec
Substituting Eq.~\eqref{eqS} in Eq.~\eqref{screen} leads to a one dimensional $q$ integral given by 
\bc
\label{FO1}
\phi_{\rm sc}({\bf r})=\sqrt{\frac{8\pi}{r}}\int_0^\infty dq~q^{1/2} \frac{\phi_{\rm ext}(q)}{\epsilon'(q,\theta_r)}\cos(qr-\pi/4)~. 
\ec
where $\epsilon'(q,\theta_r) \equiv \epsilon({\bf q},0)$ with $\theta_q \to \theta_r$.

The one dimensional integral of Eq.~\eqref{FO1} is now highly oscillatory around zero in the asymptotic limit of $r\rightarrow\infty$, for finite $q$. Thus using the 
Riemann-Lebesgue lemma, the asymptotic behaviour of the integral in Eq.~\eqref{FO1} is small and it is dominated by the contribution from the the non-analytic points of the integrand (the static dielectric function in our case).  The anisotropic static dielectric function of borophene appearing in Eq.~\eqref{FO1} has a discontinuity in its first derivative at $q = q_0 (\theta_q \to \theta_r)$; see Eq.~\eqref{q0} and Fig.~\ref{fig3}(a). Now 
following Ref.~[\onlinecite{thakur_2017}], Eq.~(\ref{FO1}) can be approximated as 
\bc\label{screen2}
\phi_{\rm sc}({\bf r})= A[q_0(\theta_r)]\int \delta \Pi^{\rm st}(q,\theta_r)S(q,r,\theta_r)e^{-qs} dq~,
\ec
where $\delta \Pi^{\rm st}(q,\theta_r) \equiv \Pi^{\rm st}[q_0(\theta_r)+q]-\Pi^{\rm st}[q_0(\theta_r)]$ is the increment of the polarization function in vicinity of the non-analytic point $q_0(\theta_r)$. 
In Eq.~\eqref{screen2}, $S(x,r,\theta_r)$ originates from the angular integration of Eq.~\eqref{eqS} and it is given by,
\bc
S(q,r,\theta_r)=\frac{\sqrt{8\pi}}{\sqrt{q_0r}}\cos[(q_0+q)r-\pi/4]~, 
\ec
and the function $A$ is explicitly given by
\bc
A[q_0(\theta_r)]=\frac{q_0\phi_{\rm ext}(q_0)v_{q=q_0}}{[1-v_{q=q_0}\Pi^{\rm st}(q_0,\theta_r)]^2}~.
\ec

\begin{figure}
\includegraphics[width=\linewidth]{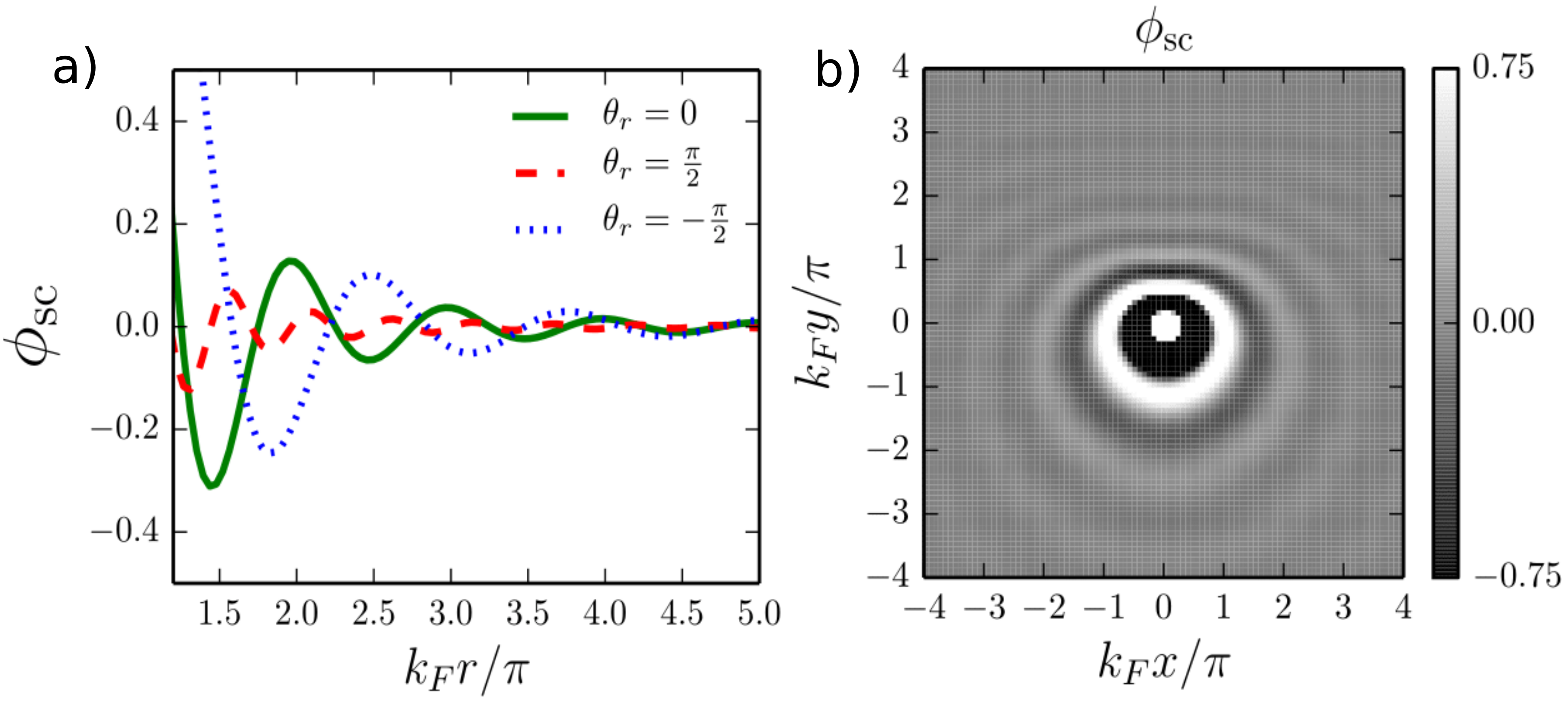}
\caption{(a) The screened impurity potential as a function of $r$, highlighting the directional dependence of the amplitude as well as the period of the Freidel oscillations in borophene. (b) Color plot of the anisotropic Friedel oscillations in the $x-y$ plane. The anisotropy in both the oscillation amplitude and period is evident. 
\label{fig4}}
\end{figure}

On evaluating Eq.~\eqref{screen2}, the asymptotic anisotropic behaviour of the screened potential is found to be similar to that in graphene: 
\bc
\phi_{\rm sc}({\bf r})=F(\theta_r) \frac{\cos(q_0 r)}{r^3}~.
\ec
Where the amplitude of Friedel oscillation $F(\theta_r)$ is given by,
\bc \label{amp}
F(\theta_r)=\frac{g A(q_0)}{\sqrt{8\pi}q_0}\frac{f(\theta_r)}{\sqrt{1-\frac{v_t^2\sin^2\theta_r}{v_x^2f^2(\theta_r)}}}\left(1-\frac{v_t\sin\theta_r}{v_xf(\theta_r)}\right)^{3/2}~.
\ec
However unlike graphene, here both $q_0$ (the period of the Friedel oscillations) and the proportionality constant (amplitude of Friedel oscillations) are direction ($\theta_r$) dependent. The directional dependence of the Friedel oscialltion amplitude and oscillation period is highlighted in Fig.~\ref{fig4}(a). Figure \ref{fig4}(b) shows the anisotropic Friedel oscillations in the $x-y$ plane. From Eq.~\eqref{amp} it can be inferred that the oscillation amplitude is maximum along the $-{\hat y}$ direction.  
The period of oscillation ($2\pi/q_0$) is also maximum along the $-{\hat y}$ direction.

\section{Effect of small static disorder on the plasmon dispersion}
\label{sec5}
In this section, we study the effect of the small static disorder on the dynamical polarization function and the plasmon mode. For dilute concentration of the impurities, the polarization function gets modified by a 
disorder averaged polarization function. Within the relaxation time approximation, the disorder averaged effective polarization is given by the Mermin formula\cite{mermin_1970,GV,amit_2015}, 
\bc\label{mermin}
\Pi_{\rm imp}^{(0)}({\bf q},\omega)=\frac{(\omega+i\tau^{-1})\Pi^{(0)}({\bf q},\omega+i/\tau)}{\omega +i\tau^{-1}\Pi^{(0)}({\bf q},\omega+i/\tau)/\Pi^{(0)}({\bf q},0)}~,
\ec
with $\tau$ being an effective relaxation time arising from the presence of the disorder. 
The dynamical long wavelength limit of the disorder averaged response function can be obtained by substituting Eq.~\eqref{long} in Eq.~\eqref{mermin}, and it is given by 
\bc\label{imp}
\Pi_{\rm imp}^{(0)}({\bf q},\omega)\approx -N_0\bigg[1-\frac{\omega'}{\sqrt{(\omega'+i/\tau)^2-q^2v_x^2f^2(\theta)}-i/\tau}\bigg]~.
\ec

Now we can easily calculate the disorder averaged plasmon dispersion in the long wavelength limit by substituting Eq.~\eqref{imp} in Eq.~\eqref{plasmon} to obtain, 
\bea
\label{imp_dispersion}
\omega_{\rm pl}^{\rm imp}({\bf q})&=& v_tq\sin\theta_q + \frac{1+v_qN_0}{1+2v_qN_0}\\
&\times & \left[\left\{v_x^2q^2f^2(\theta_q)(1+2v_qN_0)-1/\tau^2\right\}^{1/2}-i/\tau\right]. \nn
\eea
Note that $N_0 v_q = q_{\rm TF}/q$ for the Coulomb potential. The imaginary part of the plasmon dispersion is given by $\tau^{-1} \times (1+v_qN_0)(1+2v_qN_0)$ which has a finite value of $\tau^{-1}/2$ even in the $q \to 0$ limit. 
In the limit of vanishing disorder {\it i.e.}, $1/\tau \to 0$ Eq.~\eqref{imp_dispersion} reduces to Eq.~\eqref{dispersion}. As expected from the case of 
2DEG and graphene \cite{quinn_1984,amit_2015}, here also the  collective density excitations can only exist beyond a certain critical wavevector $q_c (\theta_q)$ given by 
\bc
q_c(\theta_q) = - q_{\rm TF} + \sqrt{q_{\rm TF}^2 + 1/(v_x^2 f^2(\theta_q) \tau^2)}~.
\ec
The existence of the critical wavevector, is a consequence of the fact that the collective excitations cannot exist for length scales larger than the disorder length scale (effective mean free path).

\begin{figure}[t]
\includegraphics[width=\linewidth]{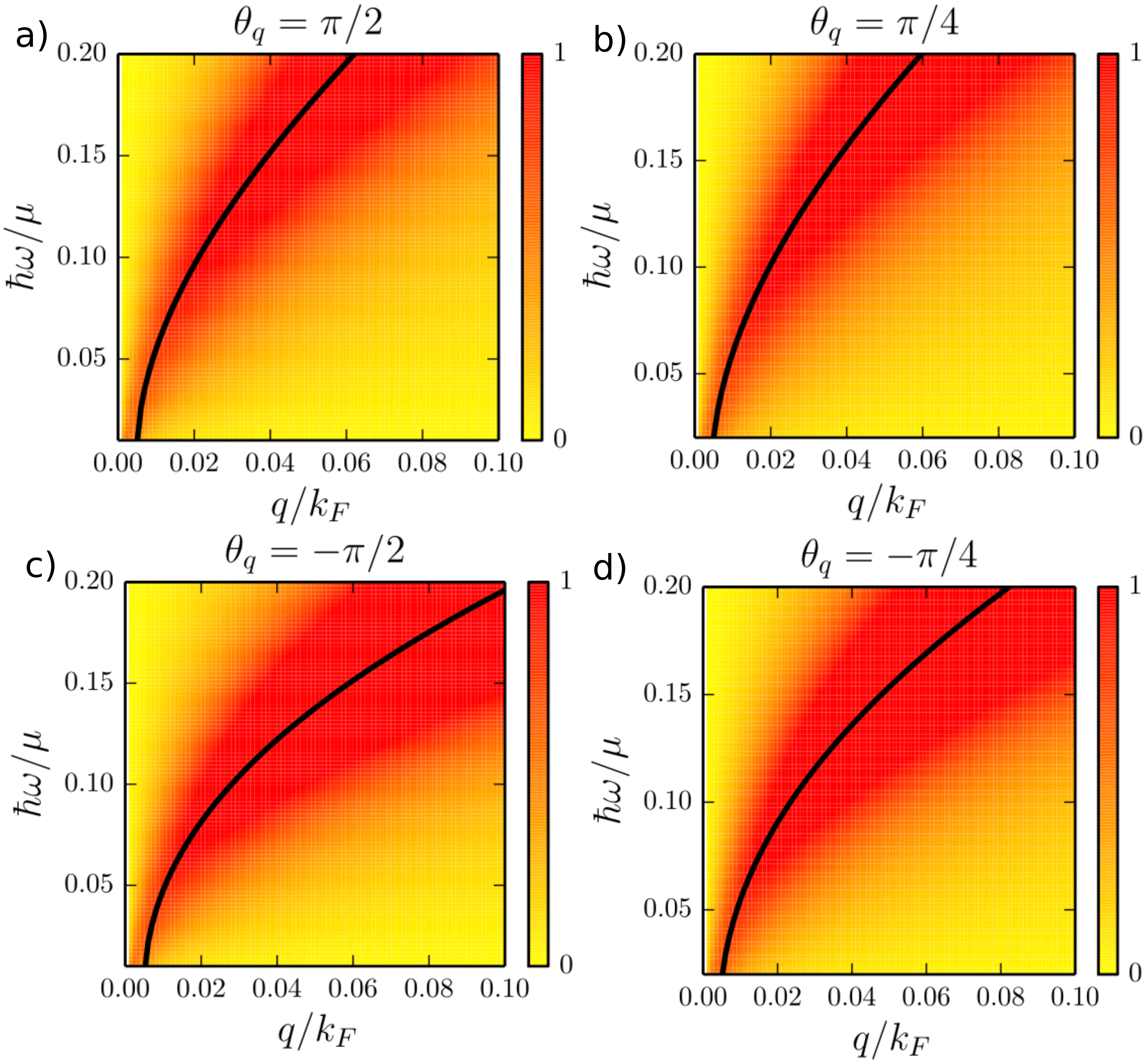}
\caption{The anisotropic loss function for the case of disordered Borophene (using Eq.~\eqref{mermin}) along with the analytical expression for the real part of the plasmon dispersion (solid black) - Eq.~\eqref{imp_dispersion} - for (a) $\theta_q = \pi/2$ (b) $\theta_q = \pi/4$  (c) $\theta_q = -\pi/2$ and (d) $\theta_q = -\pi/4$. We have chosen $2 \pi e^2 N_0/\kappa =q_{\rm TF} = 4/\pi$ and $\hbar/\tau = 0.1 \mu$ for all the panels. Note that the broadening (or damping) of the plasmon dispersion increases with increasing wave-vector (indicated by the red area).  
\label{fig5}} 
\end{figure}

Figure \ref{fig5} shows the effective energy loss function (calculated using Eq.~\eqref{Pi_final} in \eqref{mermin}) for $\hbar \tau^{-1} = 0.1 \mu$ in the $\omega-q$ plane for different directions, along with the long wavelength plasmon dispersion of Eq.~\eqref{imp_dispersion} in black line. The existence of the 
anisotropic critical wavevector, and the corresponding anisotropy in the plasmon dispersion for collective density excitations is evident. 

\section{Conclusion}
To summarize, we have calculated the density-density response function of $8-Pmmn$ borophene, which has a tilted and anisotropic Dirac cone. The polarization function is in turn used to calculate the anisotropic plasmon dispersion. The analytical form of the polarization function calculated in the dynamical long wavelength limit, matches reasonably well with the exact numerical results. We find that for wave-vectors along the mirror symmetry axis of borophene ($\hat y$ axis), both the inter- and intra-band particle-hole continuum are pushed up in energy. Consequently the plasmon dispersion is undamped for relatively higher energies for $q$ along the $\hat y$ axis. In case of small finite disorder, the plasmon mode ceases to exist below a critical anisotropic wave-vector, and beyond which it develops a finite damping even for small $q$. 
The static screening properties of $8-Pmmn$ borophene is found to be similar to that in graphene, with the coulomb potential having a asymptotic decay of $r^{-3}$. The Friedel oscillations in borophene also decay as $r^{-3}$ as in graphene, though they are highly anisotropic, with both the amplitude and periodicity of the Friedel oscillations being direction dependent. 
\label{sec6}

\appendix
\section{Calculation of the polarization function}
\label{PC}
To evaluate Eq.~\eqref{response_func}, we proceed in a manner similar to that in graphene, and define the 
retarded function \cite{stauber_2006}
\bea \label{eqA1}
\chi_D^\gamma({\bf q},\omega)&=&\frac{g}{4\pi^2\hbar^2 v_x v_y}\int_{E_k\leq D}d^2E_k \\ 
&& \sum_{\alpha=\pm}\frac{\alpha f^\gamma({\bf k},{\bf q})}{\hbar\omega'+\alpha(E_k-\gamma E_{k+q})+i\hbar\eta} \nn.
\eea
In Eq.~\eqref{eqA1}, $f^\gamma({\bf k},{\bf q})$ is the band overlap function given by,
\bc
f^\gamma({\bf k},{\bf q})=\frac{1}{2}\left(1+\gamma\frac{E_k+E_q \cos \phi_E}{\sqrt{E_k^2+E_q^2+2 E_k E_q \cos\phi_E}}\right),
\ec
and $E_{k+q}=\sqrt{E_k^2+E_q^2+2 E_k E_q \cos\phi_E}$. Here $\gamma=+1$ (-1) corresponds to intra (inter) band transitions and $D$ sets the general upper limit of the integration.
For $\mu=0$ the intrinsic polarization is given by 
\bc \label{eqA2}
\Pi^{(0)}_0({\bf q},\omega)=-\chi_\Lambda^-({\bf q},\omega),
\ec
where $\Lambda$ is the ultra violet band cut-off. In the limiting case of $\Lambda\rightarrow\infty$, Eqs.~\eqref{eqA1}-\eqref{eqA2} can be evaluated to yield 
\bc \label{Pi00}
\Pi^{(0)}_0({\bf q},\omega)=- \frac{i g}{16 \hbar^2 v_x v_y}~\frac{E_q^2}{\sqrt{(\hbar\omega-\hbar v_tq_y)^2-E_q^2}}~,
\ec
For the doped case there is finite chemical potential which gives an extra contribution to the polarization function, and it is given by 
\bc \label{pi1}
\Pi^{(0)}_1({\bf q},\omega)=\chi_\mu^+({\bf q},\omega) +\chi_\mu^-({\bf q},\omega)~.
\ec
Evaluation of Eq.~\eqref{pi1} lead to 
\begin{widetext}
\bea \label{Pi01}
\Pi^{(0)}_1({\bf q},\omega)&=& N_0\left[-1+\frac{1}{8 \mu}\frac{E_q^2}{\sqrt{\hbar^2\omega'^2-E_q^2}}\left\{G\left(\frac{\hbar \omega'+2\mu}{E_q}\right)  
 -\Theta\left(\frac{2\mu-\hbar \omega'}{E_q}-1\right)G\left[\left(\frac{2\mu-\hbar \omega'}{E_q}\right)-i\pi\right] \right. \right. \nn\\
& & \left. \left. -\Theta\left(\frac{\hbar \omega'-2\mu}{E_q}+1\right)G\left[\left(\frac{\hbar \omega'-2\mu}{E_q}\right)\right]\right\}\right]~.
\eea
\end{widetext}

In Eq.~\eqref{Pi01}, $\Theta(x)$ is the unit step function, $\omega' = \omega - v_t q_y$, and we have defined the function 
\bc \label{GG}
G(x)=x\sqrt{x^2-1}-\ln(x+\sqrt{x^2-1})~.
\ec
Equations~\eqref{Pi00}-\eqref{GG} are very similar to that of graphene\cite{stauber_2006,thakur_2017} are reproduce the results for graphene in the limiting case of $v_t \to 0$ and $v_x = v_y = v_{\rm F}$.


\begin{thebibliography}{}

\bibitem{Maier} S. A. Maier, {\it Plasmonics Fundamentals and Applications}, Springer, New York, 2007. 
%
\bibitem{Polini_review} A. N. Grigorenko,	M. Polini and K. S. Novoselov, \href{https://doi.org/10.1038/nphoton.2012.262}{Nature Photonics {\bf 6}, 749 (2012)}.
%
\bibitem{Stauber} T. Stauber, \href{https://doi.org/10.1088/0953-8984/26/12/123201}{J. Phys.: Condens. Matter 26, 123201 (2014)}.
%
\bibitem{GV} G. F. Giuliani and G. Vignale, {\it  Quantum Theory of the Electron Liquid}, Cambridge: Cambridge University Press, 2005.
%
\bibitem{appl} T. W. Ebbesen, C. Genet, and S. I. Bozhevolnyi, \href{http://dx.doi.org/10.1063/1.2930735}{Phys. Today 61 (5), 44 (2008)}; L. Novotny, \href{http://dx.doi.org/10.1063/PT.3.1167}{ibid. 64 (7), 47 (2011)}; M. I. Stockman, \href{http://dx.doi.org/10.1063/1.3554315}{ibid. 64 (2), 39 (2011)}.
%
\bibitem{fabian_2010} S. M. Badalyan, A. Matos-Abiague, G. Vignale, and J. Fabian \href{https://doi.org/10.1103/PhysRevB.81.205314}{\prb {\bf 81}, 205314 (2010)}.
%
\bibitem{SOC_plasmon} A. Agarwal, S. Chesi, T. Jungwirth, J. Sinova, G. Vignale, and M. Polini, \href{https://doi.org/10.1103/PhysRevB.83.115135}{Phys. Rev. B {\bf 83}, 115135 (2011)}.
%
\bibitem{Persistent} A. Agarwal, M. Polini, R. Fazio, and G. Vignale, \href{https://doi.org/10.1103/PhysRevLett.107.077004}{Phys. Rev. Lett. {\bf 107}, 077004 (2011).}
%
\bibitem{scholz_2012} A. Scholz, T. Stauber, and J. Schliemann \href{https://doi.org/10.1103/PhysRevB.86.195424}{\prb {\bf 86}, 195424 (2012)}.
%
\bibitem{stauber_2006} B. Wunsch, T. Stauber, F. Sols and F. Guinea, \href{https://doi.org/10.1088/1367-2630/8/12/318}{New Journal of Physics {\bf 8}, 318 (2006)}
%
\bibitem{sarma_2007}E. H. Hwang and S. Das Sarma, \href{http://dx.doi.org/10.1103/PhysRevB.75.205418}{\prb {\bf 75}, 205418 (2007)}.
%
\bibitem{polini_2011} S. H. Abedinpour, G. Vignale, A. Principi, M. Polini, W. K. Tse, and A. H. MacDonald \href{https://doi.org/10.1103/PhysRevB.84.045429}{\prb {\bf 84}, 045429(2011)}.
%
\bibitem{Peres} P. A. D. Goncalves and N. M. Peres, {\it An introduction to Graphene Plasmonics}, World Scientific, Singapore, 2016.
%
\bibitem{amit_2015} A. Agarwal and G. Vignale \href{https://doi.org/10.1103/PhysRevB.91.245407}{\prb {\bf 91}, 245407 (2015)}.
%
\bibitem{scholz_2013} A. Scholz, T. Stauber, and J. Schliemann, \href{https://doi.org/10.1103/PhysRevB.88.035135}{\prb {\bf 88}, 035135 (2013)}.
%
\bibitem{TI_plasmons} P. Di Pietro,	M. Ortolani,	O. Limaj,	A. Di Gaspare,	V. Giliberti,	F. Giorgianni,	M. Brahlek,	N. Bansal,	N. Koirala,	S. Oh,	P. Calvani	and S. Lupi, \href{https://doi.org/10.1038/nnano.2013.134}{Nature Nanotechnology {\bf 8}, 556 (2013)}.
%
\bibitem{pyat_2009} P. K. Pyatkovskiy \href{https://doi.org/10.1088/0953-8984/21/2/025506}{J. Phys.: Condens. Matter {\bf 21}, 025506 (2009)}.
%
\bibitem{thakur_2015} R. Sachdeva, A. Thakur, G. Vignale, and A. Agarwal \href{https://doi.org/10.1103/PhysRevB.91.205426}{\prb {\bf 91}, 205426 (2015)}.
%
\bibitem{thakur_2017} A. Thakur, R. Sachdeva and A. Agarwal \href{https://doi.org/10.1088/1361-648X/aa57bd}{J. Phys.: Condens. Matter {\bf 29}, 105701 (2017)}.
%
\bibitem{TLow1} T. Low, R. Roldan, H. Wang, F. Xia, P. Avouris, L. M. Moreno, and F. Guinea, 
\href{https://doi.org/10.1103/PhysRevLett.113.106802}{Phys. Rev. Lett. {\bf 113}, 106802 (2014)}.
%
\bibitem{Rodin_plasmon} A. S. Rodin and A. H. Castro Neto, 
\href{https://doi.org/10.1103/PhysRevB.91.075422}{Phys. Rev. B {\bf 91}, 075422 (2015)}.
%
\bibitem{Barun_plasmon} B. Ghosh, P. Kumar, A. Thakur, Y. S. Chauhan, S. Bhowmick, and A. Agarwal,  \href{https://arxiv.org/abs/1703.07696}{arXiv:1703.07696}. 
%
\bibitem{Poly1} Z. Zhang, E. S. Penev and B. I. Yakobson, 
\href{https://doi.org/10.1038/nchem.2521}{Nature Chemistry 8, 525 (2016)}.
%
\bibitem{Poly2} E. S. Penev, S. Bhowmick, A. Sadrzadeh, and B. I. Yakobson, \href{https://doi.org/10.1021/nl3004754}{Nano Lett. {\bf 12}, 2441 (2012)}.
%
\bibitem{Feng} B. Feng,	J. Zhang,	Q. Zhong,	W. Li,	S. Li,	H. Li,	P. Cheng,	S. Meng, L. Chen and K. Wu, 
\href{https://doi.org/10.1038/nchem.2491}{Nature Chemistry {\bf 8}, 563 (2016)}. 
%
\bibitem{Mannix}A. J. Mannix et. al., 
\href{https://doi.org/10.1126/science.aad1080}{Science {\bf 350}, 1513 (2015)}.
%
\bibitem{zabo_2016} A. D. Zabolotskiy and Yu. E. Lozovik \href{https://doi.org/10.1103/PhysRevB.94.165403}{\prb {\bf 94}, 165403(2016)}.
%
\bibitem{Zhou1} X. F. Zhou, X. Dong, A. R. Oganov, Q. Zhu, Y. Tian, and H. T. Wang, \href{https://doi.org/10.1103/PhysRevLett.112.085502}{\prl {\bf 112}, 085502 (2014).}
%
\bibitem{Feng2} B. Feng et. al, \href{https://doi.org/10.1103/PhysRevLett.118.096401}{\prl {\bf 118}, 096401 (2017).}
%
\bibitem{lopez_2016} A. Lopez-Bezanilla and P. B. Littlewood \href{https://doi.org/10.1103/PhysRevB.93.241405}{\prb {\bf 93}, 241405(R)(2016)}.
%
\bibitem{mermin_1970} N. D. Mermin \href{https://doi.org/10.1103/PhysRevB.1.2362}{\prb {\bf 1}, 2362 (1970)}
%
\bibitem{simion_2005} G. E. Simion and G. F. Giuliani \href{https://doi.org/10.1103/PhysRevB.72.045127}{\prb {\bf 72}, 045127 (2005)}
%
\bibitem{quinn_1984}G. F. Giuliani and J. J. Quinn \href{https://doi.org/10.1103/PhysRevB.29.2321}{\prb {\bf 29}, 2321(R) (1984)}
%
\end{thebibliography}
\end{document}